\documentclass[12pt]{article}
\usepackage{epsfig,amsfonts,amssymb}
\usepackage{cite}
\input epsf.sty
\usepackage{cite}

\def\ZZZ{{\hbox{ Z\kern-1.6mm Z}}}
\def\RRR{{\hbox{ R\kern-2.4mm R}}}
\def\CCC{{\hbox{ C\kern-2.0mm C}}}
\def\zzz{{\hbox{z\kern-1mm z}}}

\newcommand{\qeq}{{\hbox{=\kern-2.3mm ? \kern.5mm }}}
\renewcommand{\qeq}{=}

\newcommand{\NN}{{\cal N}}

\newcommand{\be}{\begin{equation}}
\newcommand{\ee}{\end{equation}}
\newcommand{\ben}{\begin{eqnarray}\displaystyle}
\newcommand{\een}{\end{eqnarray}}

\newcommand{\bea}[1]{\begin{eqnarray}\label{#1} }
\newcommand{\eea}{\end{eqnarray}}

\newcommand{\refb}[1]{(\ref{#1})}

\def\one{{\hbox{ 1\kern-.8mm l}}}
\def\zero{{\hbox{ 0\kern-1.5mm 0}}}

\begin{document}

\begin{center}
{\Large \bf
S-duality Action on Discrete T-duality Invariants}

\end{center}

\vskip .6cm
\medskip

\vspace*{4.0ex}

\centerline{\large \rm  Shamik Banerjee and Ashoke Sen}

\vspace*{4.0ex}

\centerline{\large \rm }

\vspace*{4.0ex}

\centerline{\large \it Harish-Chandra Research Institute}

\centerline{\large \it  Chhatnag Road, Jhusi,
Allahabad 211019, INDIA}

\vspace*{1.0ex}
\centerline{E-mail:  bshamik, sen@mri.ernet.in}

\vspace*{5.0ex}

\centerline{\bf Abstract} \bigskip

In heterotic string theory compactified on $T^6$, 
the T-duality orbits of dyons of charge $(Q,P)$
are characterized by $O(6,22;\RRR)$ invariants
$Q^2$, $P^2$ and $Q\cdot P$ together with
a set of
invariants of the discrete T-duality group $O(6,22;\ZZZ)$. 
We study the action of S-duality
group on the discrete T-duality invariants
and study its consequence
for the dyon degeneracy formula. In particular we find that
for dyons with torsion $r$, the degeneracy formula, expressed
as a function of $Q^2$, $P^2$ and $Q\cdot P$,
is required to be 
manifestly invariant under only a subgroup of the S-duality
group. This subgroup is isomorphic to $\Gamma^0(r)$.
Our analysis also shows that for a given torsion $r$, all other
discrete T-duality invariants are characterized by the
elements
of the coset $SL(2,\ZZZ)/\Gamma^0(r)$.

\vfill \eject

\baselineskip=18pt


Dyons in heterotic string theory on $T^6$ are characterized by
a pair of charge vectors $(Q,P)$ each taking value on the Narain
lattice $\Lambda$\cite{narain,nsw}. 
Given two pairs of charge vectors, an
interesting question is: under what condition can they be related
via a T-duality transformation? This question was answered in
\cite{0712.0043} where a 
complete set of T-duality invariants classifying
a pair of charge vectors $(Q,P)$ were constructed. These include
the invariants of the continuous T-duality group $O(6,22;\RRR)$
\be \label{ec1}
Q^2, \quad P^2, \quad Q\cdot P\, ,
\ee
together with a set of invariants of the discrete T-duality group
$O(6,22;\ZZZ)$. These
are defined as follows. We shall assume that the
dyon is primitive so that $(Q,P)$ cannot be written as an
integer multiple of $(Q_0,P_0)$ with $Q_0,P_0\in\Lambda$, 
but we shall not assume that $Q$ and $P$ themselves
are primitive. Now consider the intersection of the two
dimensional vector space spanned by $(Q,P)$ 
with the Narain lattice $\Lambda$. The result is
a two dimensional lattice $\Lambda_0$. Let $(e_1, e_2)$ be a
pair of basis elements whose integer linear
combinations generate this lattice. We can
always choose $(e_1,e_2)$ such that in this basis
\ben \label{ec2}
&Q = r_1 e_1, \qquad P = r_2(u_1 e_1 + r_3 e_2), \qquad
r_1,r_2,r_3,u_1\in \ZZZ^+, \nonumber \\
&\hbox{gcd}(r_1,r_2)=1,
\quad \hbox{gcd}(u_1, r_3)=1, \quad 1\le u_1\le r_3\, .
\een
It was found in \cite{0712.0043} that besides $Q^2$, $P^2$
and $Q\cdot P$, the integers $r_1$, $r_2$, $r_3$ and  $u_1$
are T-duality invariants.  Furthermore it was found that 
this is the complete set
of T-duality invariants. Thus a pair of charge vectors $(Q,P)$
can be transformed into another pair $(Q',P')$ via a T-duality
transformation if and 
only if all the invariants agree for these two
pairs.

Our first goal is to study some aspects
of the action of the S-duality transformation
\be \label{ec3}
Q\to Q' = a Q + b P, \quad P\to P'=cQ+dP, \quad
a,b,c,d\in\ZZZ, \quad ad - bc =1\, ,
\ee
on the invariants $r_1$, $r_2$, $r_3$ and $u_1$. 
Substituting \refb{ec2} into \refb{ec3}, and expressing the
resulting $(Q',P')$ as $(r_1'e_1', r_2'(u_1' e_1'+r_3' e_2'))$
for some primitive basis $(e_1',e_2')$ of $\Lambda_0$
we can determine 
$(r_1',r_2',r_3',u_1')$. Since the resuting expressions are
somewhat complicated and not very illuminating we shall not
describe them here. Instead we shall focus on some salient features
of the transformation laws of $(r_1, r_2,r_3,u_1)$. 
We first
note that the torsion $r(Q,P)$ associated with a pair of
charges $(Q,P)$, defined as\cite{askitas,0702150}
\be \label{ec4}
r(Q,P) =  Q_1 P_2 - Q_2 P_1\, ,
\ee
with $Q_i,P_i$ being the components of $Q$ and $P$ along
$e_i$, is invariant under the S-duality transformation
\refb{ec3}.
Furthermore, for
the charge vectors $(Q,P)$ given in \refb{ec2} we have
\be \label{ec5}
r(Q,P) = r_1 r_2 r_3\, .
\ee
We shall now show that one 
can always find an S-duality transformation
that brings the T-duality invariants $(r_1,r_2,r_3,u_1)$ to
$(r_1r_2r_3, 1,1,1)$ together with an appropriate
transformation on $Q^2$, $P^2$ and $Q\cdot P$ induced by
\refb{ec3}. For this we note that under the S-duality
transformation \refb{ec3}, $(Q,P)$ given in \refb{ec2}
transforms to
\be \label{ec6}
Q' = \{a r_1 + b r_2 (u_1+ k r_3)\} e_1 + b r_2 r_3 (e_2- k e_1),
\quad P' =  \{c r_1 + d r_2 (u_1+ k r_3)\} e_1 + d r_2 r_3 
(e_2- k e_1)\, ,
\ee
where $k$ is an arbitrary integer. We shall choose
\be \label{ec7}
k =\prod_i p_i\, ,
\ee
where $\{p_i\}$ represent the collection
of primes which are factors of $r_1$ but not of
$u_1$.
Now we know from
\refb{ec2} that gcd$(r_1, r_2)=1$. On the other hand 
it follows
from a result derived in appendix E of
\cite{0711.4671} that
for the
choice of $k$ given in \refb{ec7} we have
gcd$(r_1, u_1+k r_3)=1$.
Thus if we choose
\be \label{ec8}
b = r_1, \quad a = - r_2 (u_1+ k r_3)\, ,
\ee
we have gcd$(a,b)=1$ and hence we can always find $c$, $d$
satisfying $ad - bc=1$.
For this particular choice of $SL(2,\ZZZ)$
transformation we have
\be \label{ec9}
Q' = r_1 r_2 r_3 (e_2 - k e_1), \qquad
P' = - e_1 + d r_2 r_3 (e_2 - k e_1)\, .
\ee
We now define
\be \label{ec10}
e_1' = (e_2 - k e_1), \qquad
e_2' = - e_1 + (d r_2 r_3-1) (e_2 - k e_1)\, .
\ee
Since the matrix relating $(e_1',e_2')$
to $(e_1,e_2)$ has unit determinant, $(e_1',e_2')$ is a primitive
basis of the lattice $\Lambda_0$. In this basis $(Q',P')$
can be expressed as
\be \label{ec11}
Q' = r_1r_2 r_3 e_1', \qquad P' = e_1' + e_2'\, .
\ee
Comparing this with \refb{ec2} we see that for the new charge
vector $(Q',P')$ we have
\be \label{ec12}
r_1'=r_1 r_2 r_3, \quad r_2'=1, \quad r_3'=1, \quad u_1'=1\, .
\ee
This proves the desired result.

Next we shall study the subgroup of S-duality transformations
which takes a configuration with
$(r_1=r, r_2=1,r_3=1,u_1=1)$ to another configuration with
$(r_1=r, r_2=1,r_3=1,u_1=1)$. The initial configuration has
\be \label{ec13}
Q = r e_1, \qquad P = e_1 + e_2\, .
\ee
An S-duality transformation \refb{ec3} takes this to
\be \label{ec14}
Q'= a r e_1 + b (e_1 + e_2)\, , \qquad
P' = c r e_1 + d (e_1 + e_2)\, .
\ee
In order that $Q'$ is $r$ times a primitive vector, we must
demand
\be \label{ec15}
b = 0 \, \, \hbox{mod $r$}\, .
\ee
Expressing $b$ as $b_0 r$ with $b_0\in\ZZZ$ we get
\be \label{ec16}
Q'= r e_1', \qquad P' = e_1' + e_2'\, ,
\ee
where
\be \label{ec17}
e_1'=(a + b_0) e_1 + b_0 e_2, \qquad 
e_2' = (cr + d - a - b_0) e_1 + (d - b_0) e_2\, .
\ee
Since the determinant of the matrix relating $(e_1',e_2')$ to
$(e_1,e_2)$ is given by
\be \label{ec18}
 (a + b_0)(d - b_0)-b_0 (cr + d - a - b_0) =ad-bc=1\, ,
 \ee
we conclude that $(e_1',e_2')$ is a primitive basis of
$\Lambda_0$. Comparison with \refb{ec2} now shows that 
$(Q',P')$ has $r_1'=r$, $r_2'=r_3'=u_1'=1$
as required. Thus the only condition on the $SL(2,\ZZZ)$
matrix $\pmatrix{a & b\cr c & d}$ for preserving the
$(r_1=r, r_2=1,r_3=1,u_1=1)$ condition is that it must have
$b=0$ mod $r$, \i.e.\ it must be an element of $\Gamma^0(r)$.

Using this we can now determine the subgroup of $SL(2,\ZZZ)$
that takes a pair of charge vectors $(Q,P)$ with invariants
$(r_1, r_2, r_3, u_1)$ to another pair of charge vectors with
the
same invariants. 
For this we note that any $SL(2,\ZZZ)$
transformation matrix $g_0=\pmatrix{a & b\cr c & d}$ with
$a$, $b$ given in \refb{ec8} takes the set $(r_1, r_2, r_3, u_1)$
to the set $(r_1r_2 r_3, 1,1,1)$. Since the latter set is preserved by
the $\Gamma^0(r)$ subgroup of $SL(2,\ZZZ)$, the original set
must be preserved by the subgroup $g_0^{-1} \Gamma^0(r) g_0$.
This is isomorphic to the group $\Gamma^0(r)$.

To see an example of this 
consider the case 
\be \label{ecase}
r_1=r_2=1, \quad r_3=2, \quad u_1=1\, .
\ee 
In this case
the $SL(2,\ZZZ)$ transformation 
$g_0=\pmatrix{1 & 1\cr 0 & 1}$  takes
a configuration
given in \refb{ecase} to 
a configuration with $r_1=2$, $r_2=r_3=u_1=1$. Thus the
$SL(2,\ZZZ)$ transformations which take a configuration
with $(r_1=1,r_2=1, r_3=2, u_1=1)$ to a configuration with
the same discrete invariants will be of the form:
\be \label{eslnew}
\pmatrix{a' & b' \cr c' & d'} = \pmatrix{1 & -1\cr 0 & 1}
\pmatrix{ a & 2b_0\cr c & d\cr}  \pmatrix{1 & 1\cr 0 & 1}
= \pmatrix{a-c & a - c - d + 2 b_0\cr c & c+d}\, .
\ee
Since the condition $ad - 2 b_0 c=1$ requires $a$ and $d$ to be
odd, we have
\be\label{esl2}
a' + b'\in 2\ZZZ+1, \qquad c' + d'\in 2\ZZZ+1\, .
\ee
Conversely given any $SL(2,\ZZZ)$ matrix $\pmatrix{a' & b'
\cr c' & d'}$ satisfying \refb{esl2}, it can be written as
$g_0$ conjugate of the $\Gamma^0(2)$ matrix
$\pmatrix{a'+c' & -a'-c'+b'+d'\cr c' & -c' + d'}$. Thus \refb{esl2}
characterizes the subgroup of S-duality group which preserves the
condition \refb{ecase}.

The results derived so far make it clear that for a given
torsion $r$ the discrete T-duality invariants are in one to
one correspondence with the elements of the coset 
$SL(2,\ZZZ)/\Gamma^0(r)$. The representative element
for a given set of invariants
$(r_1,r_2,r_3, u_1)$ is the element $g_0^{-1}\in SL(2,\ZZZ)$
that takes a configuration with $(r_1 r_2 r_3, 1,1,1)$
to a configuration with discrete invariants $(r_1,r_2,r_3, u_1)$.
Multiplying $g_0^{-1}$ by a $\Gamma^0(r)$ element from the
right does not change the final values 
$(r_1,r_2,r_3, u_1)$ of the discrete
invariants since a $\Gamma^0(r)$ transformation does not change
the discrete T-duality invariants of the initial
configuration.

We shall now examine the consequences of these 
results for the
formula expressing the degeneracy $d(Q,P)$ -- 
or more precisely an appropriate index 
measuring the number of bosonic supermultiplets
minus the number of fermionic supermultiplets for a given set
of charges\footnote{Up to a normalization this is equal to
the helicity trace $B_6=Tr(-1)^{2h} h^6$ over all states
carrying charge quantum numbers $(Q,P)$. Here $h$ denotes
the helicity of the state.}
-- of quarter BPS dyons
as a function of $(Q,P)$. 
We note first of all that besides depending on $(Q,P)$, the degeneracy
can also depend on the asymptotic values of the moduli fields,
collectively denoted as $\phi$.
We expect
the dependence on $\phi$ to be mild, in the sense that the
degeneracy formula should be $\phi$ independent within a
given domain bounded by walls of marginal stability. 
It follows from the analysis of 
\cite{0707.1563,0707.3035} that the decays
relevant for the walls of marginal stability are of the form
\be \label{ee1}
(Q,P)\to (\alpha Q + \beta P, \gamma Q + \delta P)
+ ((1-\alpha) Q - \beta P, -\gamma Q + (1-\delta) P)\, ,
\ee
where $\alpha$, $\beta$, $\gamma$, $\delta$ are not
necessarily integers, but must be
such that $\alpha Q + \beta P$ and $\gamma Q + \delta P$
belong to the Narain lattice $\Lambda$. If we denote
by $m(Q,P;\phi)$ the BPS mass of a dyon of charge
$(Q,P)$ then the wall of marginal stability associated
with the set $(\alpha,\beta,\gamma,\delta)$ is given by
the solution to the equation
\be \label{ee2}
m(Q,P;\phi) = m(\alpha Q + \beta P, \gamma Q + \delta P;
\phi) + m((1-\alpha) Q - \beta P, -\gamma Q + (1-\delta) P;
\phi)\, .
\ee
For appropriate choice of $(\alpha,\beta,\gamma,\delta)$
this describes a codimension one subspace of
the moduli space labelled by $\phi$. Since the BPS mass
formula is invariant under a T-duality transformation
$Q\to \Omega Q$, $P\to \Omega
P$, $\phi\to \phi_\Omega$: 
\be \label{ee3}
m(\Omega 
Q,\Omega P;\phi_\Omega) = m(Q,P;\phi) \qquad
\Omega \in O(6,22;\ZZZ)
\, ,
\ee
eq.\refb{ee2} may be written as
\be \label{ee4}
m(\Omega 
Q,\Omega P;\phi_\Omega) = m(\alpha \Omega Q + \beta 
\Omega P, \gamma \Omega Q + \delta \Omega P;
\phi_\Omega) + m((1-\alpha) \Omega Q - \beta \Omega 
P, -\gamma \Omega Q + (1-\delta) \Omega P;
\phi_\Omega)\, .
\ee
This is identical to eq.\refb{ee2} with $(Q,P,\phi)$ replaced by
$(\Omega Q,\Omega P,\phi_\Omega)$.
This shows that under a T-duality transformation on charges
and moduli, the wall of marginal stability associated with
the set $(\alpha,\beta,\gamma,\delta)$ gets mapped to the wall
of marginal stability associated with the same 
$(\alpha,\beta,\gamma,\delta)$.
Thus if we consider a domain bounded by the walls of marginal
stability associated with the sets 
$(\alpha_i,\beta_i,\gamma_i,\delta_i)$ for $1\le i\le n$ -- collectively
denoted by a set of discrete variables $\vec c$ -- 
then under a simultaneous
T-duality transformation on the charges and the moduli this domain
gets mapped to a domain labelled by the same vector $\vec c$. 
The
precise shape of the domain of course 
changes since the locations of the walls in the moduli
space depends not only on 
$(\alpha_i,\beta_i,\gamma_i,\delta_i)$ for $1\le i\le n$
but also on the charges $(Q,P)$
which transform to $(\Omega Q, \Omega P)$.

We now use the fact that the dyon degeneracy formula must be
invariant under a simultaneous T-duality transformation on the
charges and the moduli, and also the fact that the dependence of
$d(Q,P;\phi)$ on the moduli $\phi$ comes only through the
domain in which $\phi$ lies, \i.e.\ the vector $\vec c$.
Since $\vec c$ remains unchanged under a T-duality transformation,
we have
\be \label{ee5}
d(Q,P; \vec c) = d(\Omega Q, \Omega P; \vec c)\, , \qquad
\Omega \in O(6,22;\ZZZ)\, .
\ee
This shows that $d(Q,P;\vec c)$ must depend only on $(Q,P)$
via the T-duality
invariants:
\be \label{ed1}
d(Q,P;\vec c) = f(Q^2,P^2,Q\cdot P, r_1, r_2, r_3, u_1; \vec c)\, ,
\ee
for some function $f$.

Let us now study the effect of S-duality transformation on this
formula. Typically an S-duality transformation will act on the
charges and hence on all the T-duality invariants and also on
the vector $\vec c$ labelling the domain bounded by the walls of
marginal stability\cite{0702141,0702150,0706.2363}. 
Indeed, as is clear from the condition \refb{ee2}, under an S-duality
transformation of the form \refb{ec3}, the wall associated
with the parameters $\pmatrix{\alpha & \beta\cr \gamma & \delta}$
gets mapped to the wall associated with
\be \label{ed1.5}
\pmatrix{\alpha' & \beta'\cr \gamma' & \delta'}
= \pmatrix{a & b\cr c & d}
\pmatrix{\alpha & \beta\cr \gamma & \delta}  
\pmatrix{a & b\cr c & d}^{-1}
\, .
\ee
Thus S-duality invariance
of the degeneracy formula now gives
\be \label{ed2}
f(Q^2,P^2,Q\cdot P, r_1, r_2, r_3, u_1; \vec c)
= f(Q^{\prime 2},P^{\prime 2},Q'\cdot P', r'_1, r'_2, r'_3, u'_1; 
\vec c\, ')\, ,
\ee
where $\vec c\, '$ stands for the collection of the sets
$\{\alpha_i', \beta_i', \gamma_i',\delta_i'\}$ computed
according to \refb{ed1.5}.
We now use the result that there exists a special class of S-duality
transformations under which 
\be \label{ed3}
(r'_1, r'_2, r'_3, u'_1) = (r_1r_2r_3,1,1,1)\, .
\ee
Using this S-duality transformation we get
\be \label{ed4}
f(Q^2,P^2,Q\cdot P, r_1, r_2, r_3, u_1; \vec c)
= f(Q^{\prime 2},P^{\prime 2},Q'\cdot P', r_1r_2r_3,1,1,1; \vec c\, ')\, .
\ee
Thus the complete information about the spectrum of quarter BPS
dyons is contained in the set of functions
\be \label{ed5}
g(Q^2,P^2,Q\cdot P, r;\vec c) \equiv 
f(Q^2,P^2,Q\cdot P, r,1,1,1;\vec c)\, .
\ee
We shall focus our attention on this function during the rest
of our analysis. Using the fact that $\Gamma^0(r)$ transformations
leave the set $(r_1=r, r_2=1, r_3=1, u_1=1)$ fixed, we see that
\be \label{ed6}
g(Q^2,P^2,Q\cdot P, r;\vec c) = 
g(Q^{\prime 2},P^{\prime 2},Q'\cdot P', r;\vec c\, ') \quad
\hbox{for} \pmatrix{Q'\cr P'} 
= \pmatrix{a & b\cr c & d} \pmatrix{Q\cr P}, \quad
\pmatrix{a & b\cr c & d}\in \Gamma^0(r)\, .
\ee
In other words, the function $g(Q^2,P^2,Q\cdot P, r;\vec c)$ is
expected to have manifest invariance under the
$\Gamma^0(r)$ subgroup of S-duality transformations.

So far our discussion has been independent of any specific
formula for the function $g(Q^2,P^2,Q\cdot P, r;\vec c)$. For
$r=1$ dyons an explicit formula for the function $g$ has been
found in a wide class of $\NN=4$ supersymmetric 
theories\cite{9607026,0412287,0505094,0506249,0508174,
0510147,0602254,
0603066,0605210,0607155,0609109,0612011,0702141,
0702150,0705.1433,0705.3874,0706.2363,0708.1270}. 
In all the known examples the function $g$ is obtained as
a contour integral of the inverse
of an appropriate modular form of a
subgroup of $Sp(2,\ZZZ)$. In particular for heterotic string
theory on $T^6$ the modular form is the well known Igusa
cusp form of weight 10 of the full $Sp(2,\ZZZ)$ group, with
the S-duality group $SL(2,\ZZZ)$ embedded in $Sp(2,\ZZZ)$
in a specific manner. Furthermore the dependence on the domain
labelled by $\vec c$ is encoded fully in the choice of the
integration contour and not in the integrand. If a similar
formula exists for $g(Q^2,P^2,Q\cdot P, r;\vec c)$ for
$r>1$, then our analysis would suggest
that the integrand should involve a modular form of a subgroup
of $Sp(2,\ZZZ)$ that contains $\Gamma^0(r)$ in the same way
that the full $Sp(2,\ZZZ)$ contains $SL(2,\ZZZ)$. It remains to be
seen if this constraint together with other physical constraints
reviewed in \cite{0708.1270} can fix the form of the integrand.

\bigskip

{\bf Acknowledgement:} We wish to thank Sandip
Trivedi for useful 
discussions.


\end{document}